\documentclass[a4paper,10pt,twoside]{cpc-hepnp}
\usepackage{multicol}
\usepackage{graphicx}
\usepackage{booktabs}
\usepackage{amssymb,bm,mathrsfs,bbm,amscd}
\usepackage{lastpage}
\usepackage{CJK}
\usepackage{verbatim}
\usepackage[tbtags]{amsmath}
\usepackage{multirow}
\usepackage{footmisc}
\usepackage{footmisc}

\newcommand{\bc}{\textbf{c}}
\newcommand{\bG}{\textbf{G}}
\newcommand{\bV}{\textbf{V}}
\newcommand{\bE}{\textbf{E}}
\newcommand{\bF}{\textbf{F}}
\newcommand{\bn}{\textbf{n}}
\newcommand{\bbb}{\textbf{b}}
\newcommand{\by}{\textbf{y}}
\newcommand{\bt}{\textbf{t}}
\newcommand{\bbm}{\textbf{m}}
\newcommand{\Eta}{\boldsymbol{\eta}}
\newcommand{\Alpha}{\boldsymbol{\alpha}}
\newcommand{\Beta}{\boldsymbol{\beta}}
\newcommand{\LLambda}{\boldsymbol{\lambda}}
\newcommand{\grad}{\ensuremath{^{\circ}}}

\begin{document}

\fancyhead[co]{\footnotesize GUAN Yinghui~ et al: Simultaneous least squares fitter based on the Lagrange multiplier method}

\footnotetext[0]{Received \today}

\title{Simultaneous least squares fitter based on the Lagrange multiplier method\thanks {Supported by Ministry of Science and Technology of China(2009CB825200), Joint Funds of National Natural Science Foundation of China(11079008), Natural Science Foundation of China(11275266) and SRF for ROCS of SEM}}

\author{%
      GUAN Yinghui$^{1}$
\quad LU Xiao-Rui $^{1;1)}$\email{xiaorui@ucas.ac.cn}\\%
\quad ZHENG Yangheng$^{1}$ %
\quad ZHU Yong-Sheng$^{1,2}$
}
\maketitle

\address{%
$^1$ University of Chinese Academy of Sciences, Beijing 100049, China\\
$^2$ Institute of High Energy Physics, Chinese Academy of Sciences, Beijing 100049, China
}

\begin{abstract}
We developed a least squares fitter used for extracting expected physics parameters from the correlated experimental data in high energy physics.
This fitter considers the correlations among the observables and handles the nonlinearity using linearization during the $\chi^2$ minimization.
This method can naturally be extended to the analysis with external inputs.
By incorporating with Lagrange multipliers, the fitter includes constraints among the measured observables and the parameters of interest.
We applied this fitter to the study of the $D^{0}-\bar{D}^{0}$ mixing parameters as the test-bed based on MC simulation. The test results
show that the fitter gives unbiased estimators with correct uncertainties and the approach is credible.
\end{abstract}



\begin{multicols}{2}
\renewcommand{\thefootnote}{\arabic{footnote}}
\section{Introduction}
It frequently happens that one wants to determine the unknown parameters from a set of correlated experimental measurements. Least squares fit~\cite{1} is an effective and standard approach for this purpose. The most general situation is the estimation problem involving the observables and unknown parameters, which are connected through a set of linear and nonlinear constraints. It is well known that if the constraints are linear equations, least squares fit gives unbiased results with correct uncertainties. For nonlinear constraints, minimization becomes more complex and linearization are often introduced so that it can be solved by linear solutions. However, those results from linearization can be slightly biased in general. Thus, good approximation in the linearization is required.

For data analysis in high energy physics experiments, the observables are mostly number of events and their relations with the parameters of interest are nonlinear in most cases. Furthermore, global fit is an important method to better constrain the parameters by combining the experimental measurements and the external inputs. In this paper, we develop an approach based on least squares fit and Lagrange multiplier method for these cases.
The statistical and systematic uncertainties of the indirect observables and their dependencies on the fit parameters~\cite{2} are considered in constructing the characteristic $\chi^2$ and the minimization procedure.

\section{Formalism}
Throughout this paper, the lowercase bold letter refers to vector quantity, the uppercase letter represents matrix quantity, the symbol \textbf{V} stands for covariance matrix.

\subsection{Construction of $\chi^{2}$}
In least squares fit with constraints, the unknown parameters \textbf{m} can be obtained by minimizing $\chi^{2}$. Referring to Ref~\cite{3,4}, we construct the $\chi^{2}$ in an extended form:
\begin{equation}
\begin{aligned}
\chi^{2}& \equiv  (\textbf{y}-{\boldsymbol{\eta}})^{T}\textbf{V}^{-1}_{\textbf{y}}(\textbf{y}- {\boldsymbol{\eta}})+ 2 \boldsymbol{\lambda_{\alpha}}^{T} \textbf{g} (\boldsymbol{\eta}, \textbf{m} ) & \\
&+  2 \boldsymbol{\lambda_{\beta}}^{T} \textbf{h} (\boldsymbol{\eta}), & \label{eq1}
\end{aligned}
\end{equation}
where $\textbf{y}$ is the vector of experimental observations, and $\boldsymbol{\eta}$ is the expected value of $\textbf{y}$. Generally, $\boldsymbol{\eta}$ is a function of \textbf{m} and their relationship can be expressed as \textbf{g}($\boldsymbol{\eta}$, \textbf{m})=\textbf{0}. $\textbf{h} (\boldsymbol{\eta})$ is the vector of constrain functions of $\boldsymbol{\eta}$. ${\boldsymbol\lambda_{\alpha}}$ and $\boldsymbol{\lambda_{\beta}}$ are the vectors of Lagrange multipliers. Minimizing the $\chi^{2}$ leads to find the optimized value of \textbf{m}. Typically, $\textbf{V}_\textbf{y}$ is determined from experimental measurements and is taken as a constant in $\chi^{2}$ fit. However, there are cases that $\textbf{V}_\textbf{y}$ depends on $\textbf{m}$. With different input \textbf{m}, the weight of each measurement should be altered. Otherwise, the result may be biased. In our case, we consider $\textbf{V}_\textbf{y}$ as $\textbf{V}_\textbf{y}(\textbf{m})$, and it will be updated in each iteration of the fit.

Let's discuss the usual cases of measurements in high energy physics experiments, where direct observables are the numbers of signal events \textbf{n}. Each item in \textbf{n} corresponds to the number of event candidates of a physics process. With extraction of the backgrounds, their expected values are functions of \textbf{m}, which in most cases are branching fractions. Usually, the signal events \textbf{n} may receive crossfeed contributions from other signal processes and contaminations from peaking backgrounds which are not belonging to the processes of interest. We use \textbf{b} to describe the number of these peaking backgrounds. The efficiencies-corrected yields, denoted by \textbf{c}, can be expressed as:

\begin{equation}
\textbf{c}=\textbf{E}^{-1}\textbf{s}=\textbf{E}^{-1}(\textbf{n}-\textbf{Fb}), \label{eq2}
\end{equation}
where, \textbf{E} is the signal efficiencies matrix, to describe detection efficiencies and crossfeed probabilities, \textbf{F} is background efficiencies matrix, to describe contamination rates from background to each signal process.

Assuming that there are external measurements $\textbf{t}$ that can be incorporated to constrain parameters of interest further, the $\chi^{2}$ can be built with all the measurements \textbf{c} and $\textbf{t}$ included in $\textbf{y}$:
\begin{equation}
\textbf{y}=
\begin{bmatrix}
c_1   \\
\vdots \\
t_1\\
\vdots\\
\end{bmatrix}  \label{eq3}
\end{equation}

In the case of that \textbf{g($\boldsymbol\eta$, m)} is nonlinear, Taylor expansion to the first order can be given as:
\begin{equation}
\textbf{g}(\boldsymbol{\eta}, \textbf{m}) \approx \textbf{g}(\boldsymbol{\eta}_{0},\textbf{m}_{0})+ \frac{\partial \textbf{g}}{\partial \textbf{m}}(\textbf{m} -\textbf{m} _{0}) + \frac{\partial \textbf{g}}{\partial \boldsymbol{\eta}}(\boldsymbol{\eta}-\boldsymbol{\eta}_{0}). \label{eq4}
\end{equation}
Here we assume that the deviation from point (\textbf{m}, $\boldsymbol\eta$) to ($\textbf{m}_{0}$, $\boldsymbol\eta_{0}$) should be small. The similar linearization is also applied on \textbf{h($\boldsymbol\eta$)}.

\subsection{Input variance}
To obtain unbiased fit results, proper handling of variance matrixes is required. According to Eq.(2), the uncertainties of \textbf{n, b, E, F} should be propagated to $\textbf{c}$ as:

\begin{equation}
\begin{aligned}
\textbf{V}_{\textbf{c}}
=&(\frac{\partial \bc}{\partial \bn})^{T}\bV_{\bn}\frac{\partial \bc}{\partial \bn}
+(\frac{\partial \bc}{\partial \bbb})^{T}\bV_{\bbb}\frac{\partial \bc}{\partial \bbb} &  \\
&+
\begin{pmatrix}
(\frac{\partial \bc}{\partial \bE})^{T} &  (\frac{\partial \bc}{\partial \bF})^{T}
\end{pmatrix}
\begin{pmatrix}
\bV_{\bE} & \textbf{C}_{\textbf{EF}} \\
\textbf{C}_{\textbf{EF}}^{T} & \bV_{\bF}
\end{pmatrix}
\begin{pmatrix}
\frac{\partial \bc}{\partial \bE} \\
\frac{\partial \bc}{\partial \bF}
\end{pmatrix}&
\end{aligned}, \label{eq5}
\end{equation}
where, $\textbf{V}_{\textbf{n}}$, $\bf V_{b}$, $\bf V_{E}$, $\bf V_{F}$  are the uncertainties of \bn, \textbf{b}, \textbf{E}, \textbf{F} respectively. Generally, the variances of \textbf{E} and \textbf{F} depend on uncertainties of estimating tracking efficiency, particle identification(PID) and so on. For the different processes, the uncertainties of the observables are correlated. Therefore, $\bf V_{E}$ and $\bf V_{F}$ have nonzero off-diagonal elements. Also \textbf{E} and \textbf{F} share many common correlated uncertainties. These common uncertainties are denoted by $\textbf{C}_{\textbf{EF}}$. More discussions about $\textbf{V}_{\textbf{c}}$ can be found in Ref.~\cite{2}.

In general cases, external measurements are not related to the internal measurements. Therefore, $\bf{V_{y}}$ is simplified as:
\begin{equation}
\bV_{\by}=
\begin{pmatrix}
\bV_{\bc} & 0 \\
0 & \bV_{\bt}
\end{pmatrix}, \label{eq6}
\end{equation}
$\bf{V_{t}}$ is the variance matrix of $\bt$. In the case of correlation exist between $\bc$ and $\bt$, the off-diagonal elements should be nonzero.

\subsection{Minimizing $\chi^{2}$}
There are many approaches in $\chi^{2}$ minimization. We adopt the iterative procedure. That is, the estimated values in step $k$, $\textbf{m}^k$, are used as seeds for calculating the estimators $\textbf{m}^{k+1}$ in the step $k+1$. The equation is formulated as~\cite{3,4}:

\begin{equation}
\begin{aligned}
\textbf{m}^{k+1} = & \textbf{m}^{k} - [ {\bG}_{\bbm}^{k} \textbf{S}_{4}^{-1} ({\bG}_{\bbm}^{k})^{T}]^{-1} {\bG}_{\bbm}^{k} \textbf{S}_{4}^{-1} & \\
&[\textbf{z}_{1} - (\textbf{G}_{\Eta}^{T})^{k} \textbf{V}_{\textbf{y}} (\textbf{H}_{\Eta})^{k} \textbf{S}_{2}^{-1} \textbf{z}_{2}], &
\end{aligned}
\label{eq7}
\end{equation}

where
\begin{equation}
\begin{aligned}
&(\textbf{G}_\textbf{m})_{il} \equiv \frac {\partial g_{l}}{\partial m_{i}},
(\textbf{G}_{\boldsymbol{\eta}})_{jl}  \equiv \frac{\partial g_{l}}{\partial \eta_{j}},& \\
&(\textbf{H}_\textbf{m})_{il} \equiv \frac {\partial h_{l}}{\partial m_{i}},
(\textbf{H}_{\boldsymbol{\eta}})_{jl}  \equiv \frac{\partial h_{l}}{\partial \eta_{j}},&  \label{eq8}
\end{aligned}
\end{equation}

\begin{equation}
\begin{aligned}
&\textbf{S}_{1} \equiv (\textbf{G}_{\Eta}^{T})^{k} \textbf{V}_{\textbf{y}} \textbf{G}_{\Eta}^{k},& \\
&\textbf{S}_{2} \equiv (\textbf{H}_{\Eta}^{T})^{k} \textbf{V}_{\textbf{y}} \textbf{H}_{\Eta}^{k}&  \label{eq9}
\end{aligned}
\end{equation}

\begin{equation}
\centering
\begin{aligned}
&\textbf{S}_{3} \equiv  (\textbf{G}_{\Eta}^{T})^{k} \textbf{V}_{\textbf{y}} \textbf{H}_{\Eta}^{k} \textbf{S}_{2}^{-1} (\textbf{H}_{\Eta}^{T})^{k}  \textbf{V}_{\textbf{y}}
 (\textbf{G}_{\Eta}^{k}),  & \\
& \textbf{S}_{4} \equiv \textbf{S}_{1} -  \textbf{S}_{3}, &  \label{eq10}
\end{aligned}
\end{equation}

\begin{equation}
\begin{aligned}
&\textbf{z}_{1} \equiv \textbf{g}^{k}+\textbf{G}_{\Eta}^{T}(\textbf{y}-\Eta^{k}), &  \\
& \textbf{z}_{2} \equiv \textbf{h}^{k}+\textbf{H}_{\Eta}^{T}(\textbf{y}-\Eta^{k}). &  \label{eq11}
\end{aligned}
\end{equation}
The fit procedure is to reiterate Eq.(7) until the $\chi^{2}$ converges. Then we obtain the variance matirx as£º
\begin{equation}
\centering
\textbf{V}_{\textbf{m}} = \textbf{S}_{5} \textbf{V}_{\textbf{y}} \textbf{S}_{5}^{T},   \label{eq13}
\end{equation}
where
\begin{equation}
\begin{aligned}
\textbf{S}_{5} \equiv & [ {\bG}_{\bbm} \textbf{S}_{4}^{-1} {\bG}_{\bbm}^{T}]^{-1} {\bG}_{\bbm} \textbf{S}_{4}^{-1} {\bG}_{\Eta}^{T} & \\
&[\textbf{I}-\textbf{V}_{\textbf{y}}\textbf{H}_{\Eta}\textbf{S}_{2}^{-1}\textbf{H}_{\Eta}^{T}],& \label{eq12}
\end{aligned}
\end{equation}
where \textbf{I} indicates the unit matrix. More details about deducing Eq.(7-13) are put in the Appendix A. In our specific case, \textbf{c} is dependent on \textbf{m}(through \textbf{b}). Note that $\partial \textbf{c}/ \partial \textbf{m}$ can be ignored in $\chi^{2}$ minimization, because the elements of \textbf{F} are very small in general. $\partial \textbf{V}_\textbf{c}/ \partial \textbf{m}$ is not considered in deriving Eq.(7).  This special treatment avoids the potential bias~\cite{2}, which is introduced by this item. However, in each iteration all the input variables that depend on \textbf{m} are recalculated, including $\textbf{V}_{\textbf{c}}$ and \bc.

\renewcommand{\thefootnote}{\arabic{footnote}}
\section{Monte Carlo study}
The fitter is developed based on ROOT~\cite{5} framework. We test it in the measurement of $D^{0}-\bar{D}^0$ mixing parameters by toy Monte Carlo (MC) simulation under the environment of the BESIII experiment~\cite{6}, where $D$-pair is produced through $e^{+}e^{-}\rightarrow\psi(3770)\rightarrow D\bar{D}$, and they are in a quantum-correlated $C$-odd system~\cite{7,8}. The measurement of their decay rates provide unique opportunity for measuring $D^{0}-\bar{D}^0$ mixing parameters~\cite{9,10,11}. We use ten signal processes as listed in Table~\ref{tab1}~\cite{12}.

\begin{center}
\tabcaption{Signal processes involved in the test. $f^{cor}$ are the correlated ($C$-odd) effective $D^0\bar{D^{0}}$ branching ratios, to the leading order in $x_{D}$, $y_{D}$ and $R_{WS}$, divided by the branching ratios $\mathcal{B}_{i}$ of a isolated $D$ for modes $i$ and $\mathcal{B}_{i}\mathcal{B}_{j}$ for modes $\{i,j\}$. }
\footnotesize
\begin{tabular*}{75mm}{c@{\extracolsep{\fill}}cc}
\toprule
 $D$ decay mode & $f^{cor}$     \\
\hline
$K^{-}\pi^{+}$ & $1+R_{WS}$   \\
$K^{+}K^{-}$  &  2             \\
$K_S\pi^0$  &  2              \\
\hline
$K^{-}\pi^{+},K^{+}\pi^{-}$   &  $(1+R_{WS})^2 - 4r \cos\delta_{K\pi} (r \cos\delta_{K\pi}+ y_{D})$      \\
$K^{-}\pi^{+}, K^{+}K^{-}$          &  $1+R_{WS} + 2r \cos\delta_{K\pi} + y_{D}$      \\
$K^{-}\pi^{+}, K_S\pi^0$   &  $1+R_{WS} - 2r \cos\delta_{K\pi} - y_{D}$     \\
$K^{-}\pi^{+}, K^{+}e^-\bar{\nu}_e$    &  $1-ry_{D}\cos\delta_{K\pi}-rx_{D}\sin\delta_{K\pi}$         \\
$K^{+}K^{-}, K_S\pi^0$             &  4   \\
$K^{+}K^{-}, Ke\nu_e$             &  2$(1+y_{D})$   \\
$K_S\pi^0, Ke\nu_e$             &  2$(1-y_{D})$   \\
\bottomrule
\end{tabular*}
\label{tab1}
\end{center}

The fit is expected to reproduce nine parameters: $N_{DD}$, $\mathcal{B}$($K\pi$), $\mathcal{B}$($KK$), $\mathcal{B}$($K_{S}\pi^0$), $\mathcal{B}$($Ke\nu$), $r$, $\delta_{K\pi}$, $x_{D}$, $y_{D}$. $N_{DD}$ indicates the total number of produced $D^{0}\bar{D}^{0}$ pairs; $\mathcal{B}$ indicates the branching ratios; $-\delta_{K\pi}$ is the relative phase between the doubly Cabibbo-suppressed $D^{0}\rightarrow K^{+}\pi^{-}$ amplitude and the corresponding Cabibbo-favored $\bar{D}^{0}\rightarrow K^{+}\pi^{-}$ amplitude: $ <K^{+}\pi^{-}|D^{0}>/<K^{+}\pi^{-}|\bar{D}^{0}> \equiv r e^{-i\delta_{K\pi}}$; $x_{D}$, $y_{D}$ are parameters describes charm mixing, for the details of these definitions, we refer to Ref~\cite{12}. We input $N_{DD}=5.0 \times 10^{6} $, which roughly corresponds to those yields in $3.0 fb^{-1}$ data of $e^{+}e^{-}\to D\bar{D}$ at the $\psi(3770)$ resonance. The values of other input parameters are taken as the world-average values~\cite{13} with Gaussian smearing. The width of Gaussian is taken as the error of the corresponding parameter. Detection efficiencies for these processes are determined from MC sample of simulating the BESIII detector. We assume 0.5\% peaking backgrounds (from $\rho\pi$ processes) for the modes involved with $D \rightarrow K_{S}\pi^0$. We apply correlated systematic uncertainties of 1\% for tracking efficiencies, 2\% for $\pi^{0}$ finding and 4\% for $K_{S}$ finding. All the event yields are fluctuated according to Poisson statistics.
In the fit to the MC sample, we take inputs of data from other experimental measurements, which can provide more constraints on parameters of interest. There are seven external inputs in the test: $R_{WS}$, $r^{2}$, $\delta_{K\pi}$, $x_{D}$, $y_{D}$, $x'^2$ and $y'$ and their uncertainties are assumed to be uncorrelated.
Elements of {\bf{c}, \bf{t}, \bf{m}} and the constrain functions which are used in the MC test are listed in Table~\ref{tab2}\footnote{Relationships between $c_{i}$ and $\bf{m}$ could be found in Table~\ref{tab1}.}.
\begin{center}
\tabcaption{Elements of \textbf{c}, \textbf{t}, \textbf{m} and constrain functions used in the MC test. Each element of the  \textbf{c} indicates the efficiencies-corrected yield corresponding each process listed in Table~\ref{tab1}.}
\footnotesize
\begin{tabular*}{75mm}{c@{\extracolsep{\fill}}cccc}
\toprule
\textbf{c}  & \textbf{t}   & \textbf{m}  &  Relationship\\
\hline
\multirow{7}*{$c_{i}$}\  &  $R_{WS}$          &                   &   \\
                        &  $r^{2}$          &  $N_{DD}$         &
                         $R_{WS} =  r^2 + r y_{D}\cos(\delta_{K\pi})$ \\
& $\delta_{K\pi}$   & $\mathcal{B}(KK)$     &  $-rx_{D}\sin(\delta_{K\pi})+\frac{(x_{D}^2+y_{D}^2)}{2},$  \\
&  $x_{D}$          & $\mathcal{B}(K_{S}\pi^0)$ &$x'=  x_D \cos\delta_{K\pi}+y_{D}\sin\delta_{K\pi},$ \\
&  $y_{D}$          & $\mathcal{B}(K\pi)$  &  $y'=  y_D \cos\delta_{K\pi} - x_{D} \sin\delta_{K\pi}.$ \\
&  $x'^2$           & $\mathcal{B}(Ke\nu)$ & \\
&  $y'$             & $r$  &  \\
\bottomrule
\end{tabular*}
\label{tab2}
\end{center}
We do ten thousands times of sampling and perform the least squares fit for each sample. The pull distributions for nine fit parameters are shown in Fig.~\ref{fig1}. All the pull distributions agree well with the normal distribution and the confidence level is flat. This indicates that the fitter provides unbiased estimations of the parameters of interest and good convergence. Slight asymmetries in pull distributions may present, due to the nonlinearity. Table~\ref{tab3} lists the correlation coefficients among the fit parameters. As we expect, branching fractions tend to be positively correlated with each other and negatively correlated with $N_{DD}$.

\begin{center}
\includegraphics[width=6.5cm]{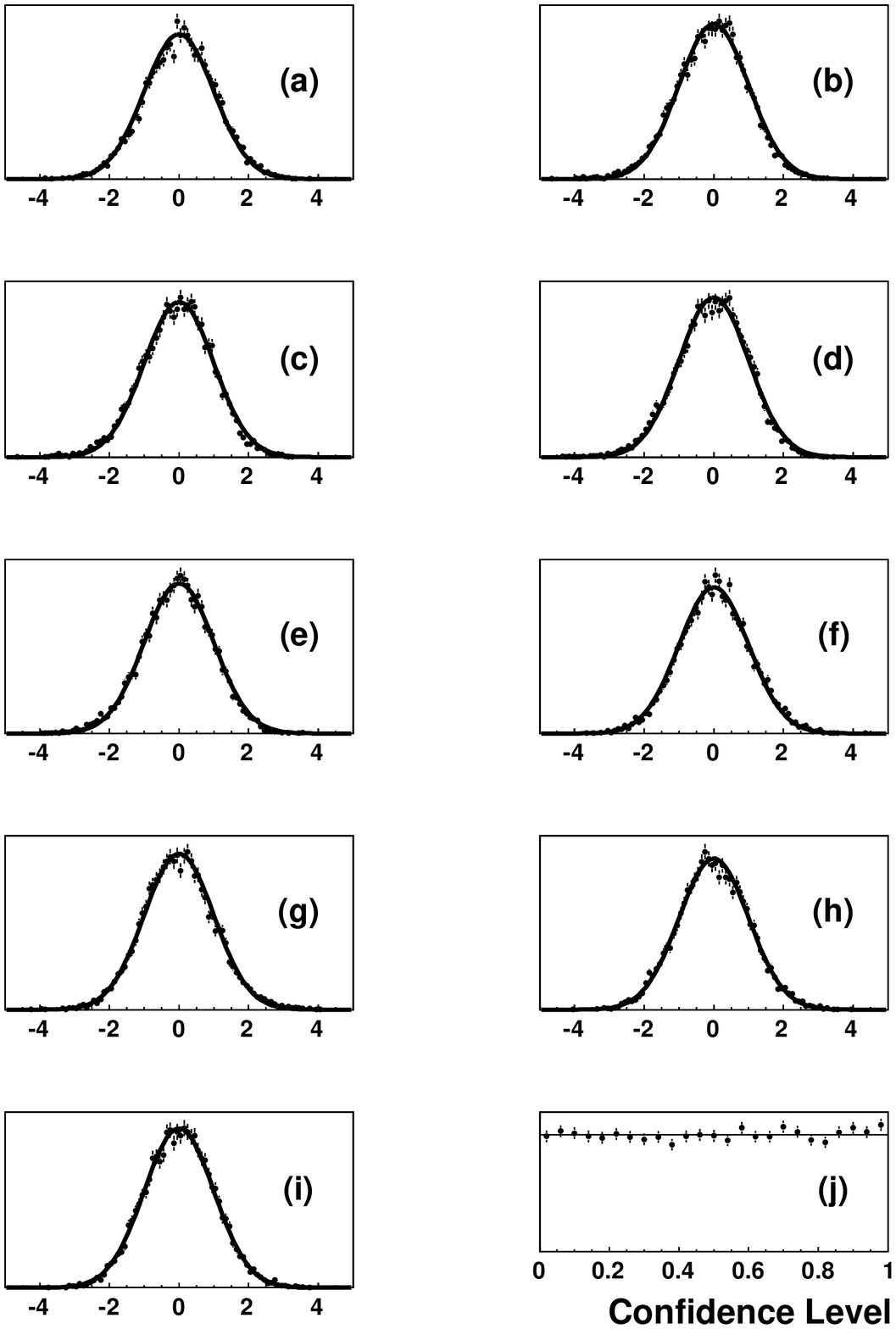}
\figcaption{Pull distributions of $N_{DD}$(a), $\mathcal{B}(K\pi$)(b), $\mathcal{B}(KK$)(c), $\mathcal{B}(K_{S}\pi^0$)(d), $\mathcal{B}(Ke\nu$)(e), $r$(f), $\delta_{K\pi}$(g), $x_{D}$(h), $y_{D}$(i) overlaid with normal distributions and the confidence level distribution(j) overlaid with a line with zero slope.}
\label{fig1}
\end{center}

We also estimate the sensitivity of measuring $y_{D}$ and $\delta_{K\pi}$ under the current statistics.
Considering more available modes, in this estimation, events yields for $CP$ eigenstates and semi-leptonic processes are scaled by a factor of 2 roughly. We input world-average $\delta_{K\pi}=22.1^{+9.7}_{-11.1}(\grad)$ and $y_{D}=0.75\pm0.12(\%)$~\cite{13} for the fit test. One-dimensional confidence curves of the fit of $y_{D}$ and $\delta_{K\pi}$ are shown in Fig.~\ref{fig2}. The curves are obtained by repeating the fits at fixed value of $y_{D}$ or $\delta_{K\pi}$ in one MC trial and recording the change from the minimum $\chi^{2}_{\rm min}$. The uncertainties of output $\delta_{K\pi}$ and $y_{D}$ are determined to be $^{+8.3}_{-9.4}(\grad)$~\footnote{The two uncertainties are evaluated using two values of asymmetric uncertainty of input $\delta_{K\pi}$ respectively.} and 0.10\% respectively. The results show that uncertainties on $y_{D}$ and $\delta_{K\pi}$ are both improved by about 15\%.

\begin{center}
\includegraphics[width=7.0cm]{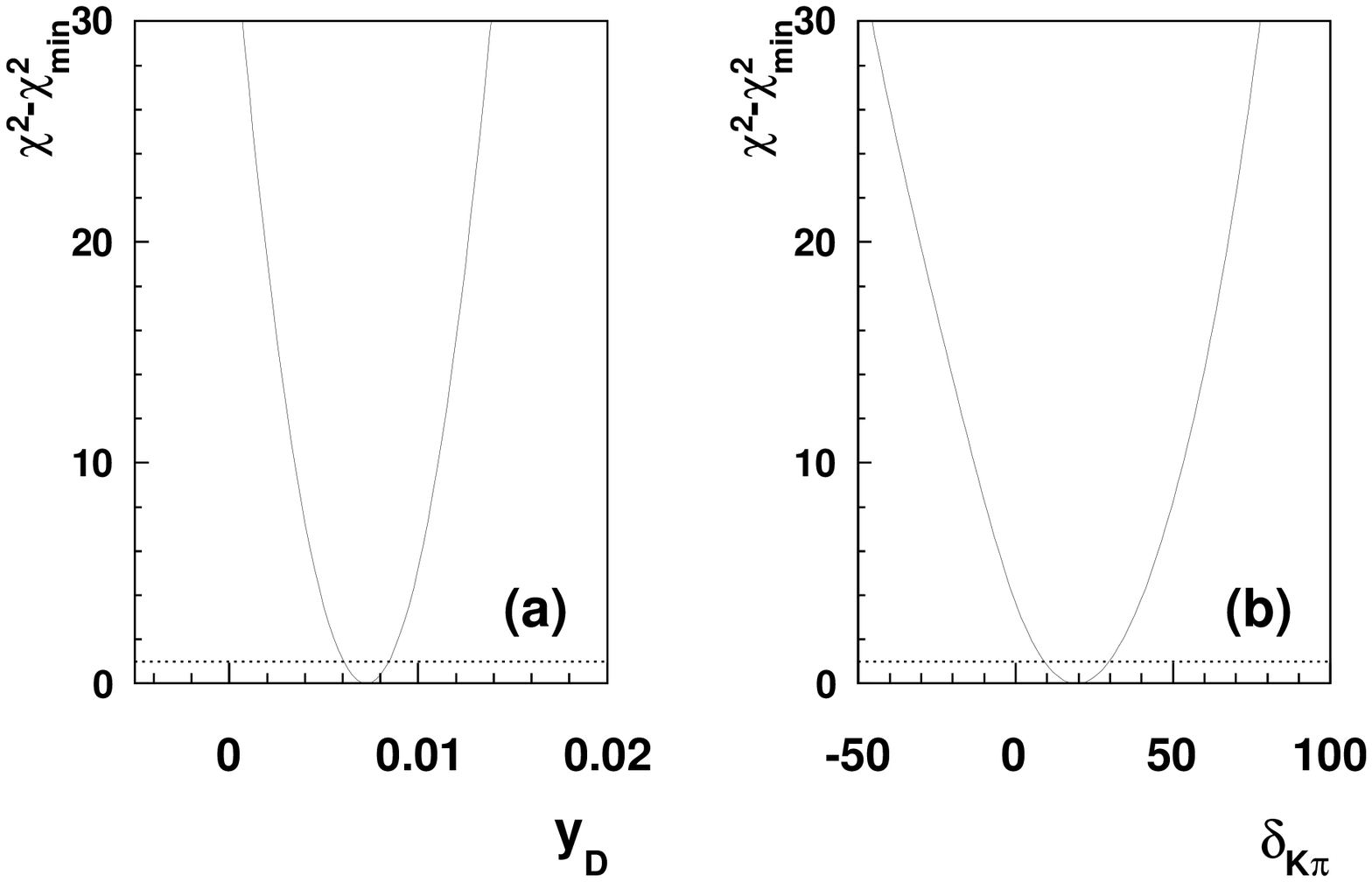}
\figcaption{The function $\Delta\chi^{2}$=$\chi^{2}$-$\chi^{2}_{\rm min}$ for $y_{D}$(a) and $\delta_{K\pi}$(b). The dashed line denotes the points where $\Delta\chi^{2}$=1.}
\label{fig2}
\end{center}

\ruleup
\end{multicols}

\begin{center}
\tabcaption{Correlation coefficients, including systematic uncertainties, for the parameters determined by the fit with MC samples.}
\footnotesize
\begin{tabular*}{130mm}{c@{\extracolsep{\fill}}cccccccccc}
\toprule
& $N_{DD}$ & $\mathcal{B}(K\pi)$ & $\mathcal{B}(KK)$ &  $\mathcal{B}(K_{S}\pi^0)$ &  $\mathcal{B}(Ke\nu)$ &  $r$ &  $\delta_{K\pi}$ &  $x_{D}$ & $y_{D}$  \\
$N_{DD}$   & 1 & -0.63 & -0.63 & -0.24& -0.09 & -0.02 & 0.03 & -0.01 &  -0.01 \\
$\mathcal{B}(K\pi)$ &   & 1 & 0.96 & 0.15 & 0.65 &  0.01 & -0.02 & 0.00 & 0.01  \\
$\mathcal{B}(KK)$   &   &   & 1 & 0.15 & 0.63 &  0.01 & -0.02 & 0.00 & 0.01    \\
$\mathcal{B}(K_{S}\pi^0)$ & & &  & 1 & -0.03 &  0.01  & -0.02 & 0.00 & 0.01  \\
$\mathcal{B}(Ke\nu)$  & & & & & 1&  -0.00 & 0.00 & -0.00 & 0.01   \\
$r$  & & &  & &  & 1& 0.06& 0.11 & -0.28 \\
$\delta_{K\pi}$  &  & & &&&& 1 & -0.09& 0.09  \\
$x_{D}$ &  &  & & & & & & 1 & -0.09  \\
$y_{D}$ &  & & & & & & & & 1 \\
\bottomrule
\end{tabular*}
\label{tab3}
\end{center}
\begin{multicols}{2}

\section{Summary}
We developed a least squares fitter, which extracts the expected parameters by combining the experimental measurements and the external inputs. Lagrange multiplier method is adopted accounting for constraints among the observables and the expected parameters. In the fitter, the observables and the input covariance matrix are supposed to be dependent with the expected parameters and they need to be renewed in each iteration step during minimization procedure. With correct input of the error matrix of the observables, the fitter gives unbiased estimations with correct uncertainties of the expected parameters. The test on toy MC validates the credibility of the fitter.

\end{multicols}

\vspace{5mm}
\begin{multicols}{2}
\subsection*{Appendix A}
\begin{small}
\noindent{\bf Formulas for iterative process}

\begin{subequations}
\renewcommand{\theequation}{A\arabic{equation}}

Following the similar procedure presented in Ref~\cite{3,4}, one can obtain Eq.(7-13).
By assuming the deviation of $\chi^{2}$ to $\boldsymbol\eta$, \textbf{m}, $\LLambda_{\Alpha}$, $\LLambda_{\Beta}$ equal to zero, we obtain
\begin{equation}
-2\textbf{V}_{\textbf{y}}^{-1}(\textbf{y}-\boldsymbol{\eta})+2\textbf{G}_{\boldsymbol{\eta}}\boldsymbol{\lambda_{\alpha}}
+2\textbf{H}_{\boldsymbol{\eta}}\boldsymbol{\lambda_{\beta}}=0, \label{a1}
\end{equation}
\begin{equation}
2\textbf{G}_{\textbf{m}}\boldsymbol{\lambda_{\alpha}}=0, \label{a2}
\end{equation}
\begin{equation}
2\textbf{g} (\boldsymbol{\eta}, \textbf{m} )=0,  \label{a3}
\end{equation}
\begin{equation}
2\textbf{h} (\boldsymbol{\eta})=0.  \label{a4}
\end{equation}
$\textbf{m}^{k+1}$, $\Eta^{k+1}$, $\LLambda_{\Alpha}^{k+1}$, and $\LLambda_{\Beta}^{k+1}$ are used as inputs to the next iteration. Eq.(A1) and (A2) can be re-expressed as:
\begin{equation}
\textbf{V}_{\textbf{y}}^{-1}(\Eta^{k+1}-\textbf{y})+\textbf{G}_{\Eta}^{k}\LLambda_{\Alpha}^{k+1}+\textbf{H}_{\Eta}^{k}\LLambda_{\Beta}^{k+1}=0, \label{a5}
\end{equation}
\begin{equation}
\textbf{G}_{\textbf{m}}^{k} \LLambda_{\Alpha}^{k+1}=0. \label{a6}
\end{equation}
With Taylor expansion, Eq.(A3) and (A4) become:
\begin{equation}
\textbf{g}^{k}+(\textbf{G}_{\Eta}^{T})^{k}(\Eta^{k+1}-\Eta^{k})+({\bG}_{\bbm}^{T})^{k}( \bbm^{k+1}-\bbm^{k})=0, \label{a7}
\end{equation}
\begin{equation}
\textbf{h}^{k}+(\textbf{H}_{\Eta}^{T})^{k}(\Eta^{k+1}-\Eta^{k}) =0. \label{a8}
\end{equation}
From Eq.(A5), we have
\begin{equation}
\Eta^{k+1}= \textbf{y}-\textbf{V}_{\textbf{y}}\textbf{G}_{\Eta}^{k}\LLambda_{\Alpha}^{k+1} - \textbf{V}_{\textbf{y}} \textbf{H}_{\Eta}^{k}\LLambda_{\Beta}^{k+1}.  \label{a9}
\end{equation}
With input of Eq.(A9), Eq.(A7) and Eq.(A8) are re-written as:
\begin{equation}
\begin{aligned}
&\textbf{z}_{1} -\textbf{S}_{1} \LLambda_{\Alpha}^{k+1} - (\textbf{G}_{\Eta}^{T})^{k}\textbf{V}_{\textbf{y}} \textbf{H}_{\Eta}^{k} \LLambda_{\Beta}^{k+1} & \\
&+ ({\bG}_{\bbm}^{T})^{k} (\bbm^{k+1}-\bbm^{k})=0, &
\end{aligned} \label{a10}
\end{equation}
\begin{equation}
\textbf{z}_{2}- (\textbf{H}_{\Eta}^{T})^{k} \textbf{V}_{\textbf{y}}\textbf{G}_{\Eta}^{k}\LLambda_{\Alpha}^{k+1} - \textbf{S}_{2}\LLambda_{\Beta}^{k+1} =0.  \label{a11}
\end{equation}
Then
\begin{equation}
\LLambda_{\Beta}^{k+1} = \textbf{S}_{2}^{-1}(\textbf{z}_{2} - (\textbf{H}_{\Eta}^{T})^{k} \textbf{V}_{\textbf{y}} \textbf{G}_{\Eta}^{k}\LLambda^{k+1}_{\Alpha} ). \label{a12}
\end{equation}
$\LLambda_{\Beta}^{k+1}$ in Eq.(A10) is substituted as
\begin{equation}
\begin{aligned}
&\textbf{z}_{1} -  \textbf{S}_{4} \LLambda^{k+1}_{\Alpha} -  (\textbf{G}_{\Eta}^{T})^{k} \textbf{V}_{\textbf{y}} (\textbf{H}_{\Eta})^{k} \textbf{S}_{2}^{-1} \textbf{z}_{2} & \\
&+  ({\bG}_{\bbm}^{T})^{k} ( \bbm^{k+1}-\bbm^{k}) =0 \label{a13}. &
\end{aligned}
\end{equation}
Then $\LLambda^{k+1}_{\Alpha}$ becomes
\begin{equation}
\begin{aligned}
& \LLambda^{k+1}_{\Alpha} =   \textbf{S}_{4}^{-1}[ \textbf{z}_{1} - (\textbf{G}_{\Eta}^{T})^{k} \textbf{V}_{\textbf{y}} (\textbf{H}_{\Eta})^{k} \textbf{S}_{2}^{-1} \textbf{z}_{2} & \\
&+  ({\bG}_{\bbm}^{T})^{k}( \bbm^{k+1}-\bbm^{k}) ]. &  \label{a14}
\end{aligned}
\end{equation}
Combing Eq.(A14) and Eq.(A6), we would derive out $\textbf{m}^{k+1}$ in Eq.(7). The estimators $\Eta^{k+1}$ and $\LLambda^{k+1}_{\Beta}$ are obtained from Eq.(A9) and Eq.(A12).

The variance matrixes $\textbf{V}_{\textbf{m}}$ and $\textbf{V}_{\Eta}$ and their correlated variances can be obtained from Eq.(7) and Eq.(A9):
\begin{equation}
\textbf{V}_{\textbf{m}} = ( \frac{\partial \textbf{m}} {\partial \textbf{y}})^{T} \textbf{V}_{\textbf{y}} ( \frac{\partial \textbf{m}} {\partial \textbf{y}}),
\end{equation}
\begin{equation}
\textbf{V}_{\Eta} = ( \frac{\partial \Eta} {\partial \textbf{y}})^{T} \textbf{V}_{\textbf{y}} ( \frac{\partial \Eta} {\partial \textbf{y}}),
\end{equation}
\begin{equation}
cov(\Eta,\textbf{m}) = ( \frac{\partial \Eta} {\partial \textbf{y}})^{T} \textbf{V}_{\textbf{y}} ( \frac{\partial \textbf{m}} {\partial \textbf{y}}).
\end{equation}

\end{subequations}
\end{small}
\end{multicols}

\vspace{30mm}
\vspace{-1mm}
\centerline{\rule{80mm}{0.1pt}}
\vspace{2mm}

\begin{multicols}{2}

\end{multicols}

\vspace{10mm}

\clearpage
\end{document}